\documentclass[12pt]{iopart}
\newcommand{\DD}{{D}} 
\newcommand{\RR}{\mathcal{R}} 
 
\newcommand{\Sol}{{S}} 
\newcommand{\phiout}{{\Phi^\mathrm{out}_\mathrm{eq}}} 
\newcommand{\phiin}{{\Phi^\mathrm{in}_\mathrm{eq}}} 
\newcommand{\phiinf}{\bar \Phi} 
\newcommand{\phismallout}{\phi^\mathrm{out}}

\newcommand{\ch}[1]{\textcolor{black}{{#1}}}

\DeclareMathAlphabet{\mathpzc}{OT1}{pzc}{m}{it}
\usepackage{graphicx}

\usepackage{color}

\begin{document}

\title[Droplet Ripening in Concentration Gradients]{Droplet Ripening in Concentration Gradients}

\author{Christoph A.\ Weber$^{1,2}$, Chiu Fan Lee$^3$ and Frank J\"{u}licher$^{1,2}$}
\address{$^1$ Max Planck Institute for the Physics of Complex Systems,
N\"{o}thnitzer Str.~38, 01187 Dresden,
Germany
}
\address{$^2$ Center for Advancing Electronics Dresden cfAED, Dresden, Germany}

\address{$^3$ Department of Bioengineering, Imperial College London, South Kensington Campus, London SW7 2AZ, U.K.}

\ead{c.lee@imperial.ac.uk, julicher@pks.mpg.de}
\vspace{10pt}


\begin{abstract}
Living cells use phase separation and concentration gradients to organize 
chemical compartments in space.
Here, we present a theoretical study of droplet dynamics in gradient systems. 
We derive the corresponding growth law of droplets 
and find that droplets exhibit a drift velocity and position dependent growth. 
As a consequence, the dissolution boundary moves through the system, thereby segregating droplets to one end. 
We show that 
for steep enough gradients, the ripening leads to a transient arrest \ch{of droplet growth} that is induced by an
narrowing of the droplet size distribution.
\end{abstract}

%
\vspace{2pc}
%

%
%
%

\section{Introduction: Droplet ripening in concentration gradients in biology}

Living cells have to organize many molecules in
space and time in order to build compartments which can perform 
certain biological functions. 
The formation of these compartments is often regulated by spatially
heterogenous distributions of molecular species.
An example is the polarized 
distribution of polarity proteins in the course of asymmetric cell division~\cite{brangwynne2011soft, hyman2014liquid, brangwynne2015polymer}. 
During asymmetric cell division, molecules of the cell cytoplasm are distributed unequally between both daughter cells~\cite{cowanhyman2004asymmetric, betschinger2004dare}. 
This can be studied in the first division of the fertilized egg of
the roundworm \emph{C.elegans}.  
RNA-protein aggregates called P-granules are
segregated to the posterior side of the cell and are located in the posterior daughter cell after division. P-granules are liquid like droplets that form by phase separation from the cell cytoplasm~\cite{Brangwynne_2009, brangwynne2011soft, hyman2014liquid, brangwynne2015polymer}. 
The segregation and ripening of P-granule droplets toward the posterior is driven by a concentration gradient of the protein Mex-5 that regulates droplet dynamics~\cite{Brangwynne_2009, Lee_2013, saha2016polar}.

\begin{figure}[t]
\centering
\includegraphics[width=0.8\textwidth]{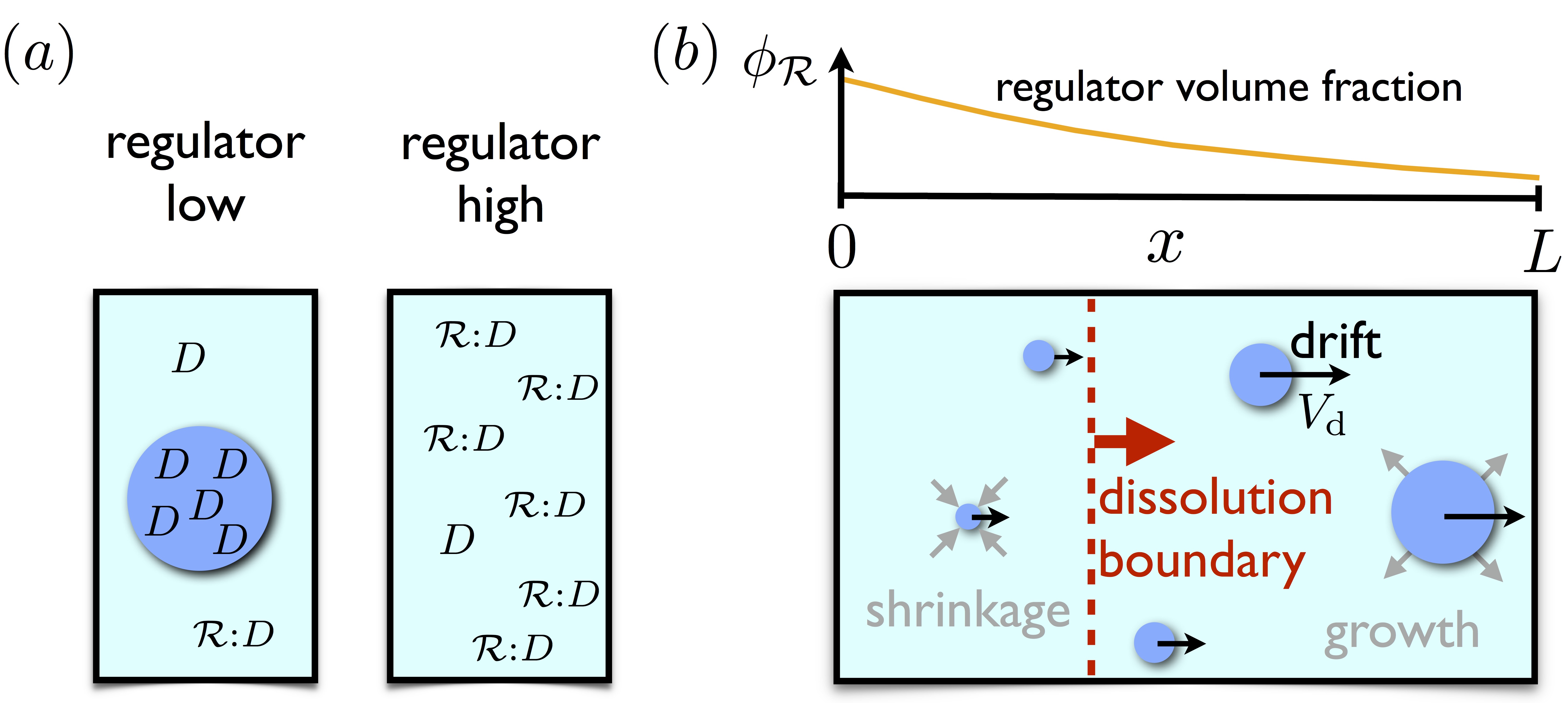} 
\caption{\label{fig:Fig_0}
(a) 
Schematic representation of the regulation of droplet (blue dot)  formation by a regulator $R$ which can bind to droplet material $D$ to form the product $\RR$:$\DD$. 
(b) Illustration of droplet ripening in a gradient of regulator volume fraction
$\phi_R$ (orange).
At each time point a boundary (red dashed) divides the system into domains of growth and shrinkage. 
This boundary moves to the right (red arrow) leaving a region of dissolving drops behind. 
Droplets drift (black arrows) with velocity $V_\mathrm{d}$.}
\end{figure}

The ripening of drops guided by a concentration gradient 
of molecules that regulate phase separation fundamentally differs  from classical Ostwald-ripening.
In the case of Ostwald ripening, droplets are uniformly distributed throughout the system and the droplet size distribution broadens with time~\ch{\cite{Lifshitz_Slyozov_61,wagner61,yao1993theory, Bray_Review_1994}}.
If a concentration gradient of a regulator 
component is maintained, for example by sources and sinks~\cite{saha2016polar}, or {\it via} position-dependent reaction kinetics~\cite{tenlen08, griffin11},
there is a broken symmetry generating a bias of droplet positions.
Recently, droplet segregation in a concentration gradient has been discussed using a simplified model~\cite{Lee_2013}.
However, the dynamics of 
droplets ripening
in a gradient of regulating molecules has not been explored (figure~\ref{fig:Fig_0}(a)).

In this paper we present a theoretical study of droplet ripening in a concentration
gradient of a regulator that affects phase separation.
Considering a simplified theory we extract generic 
physical features of droplet growth in the presence of concentration gradients.
The generic features we study here are the spatially dependent, local equilibrium concentration and a spatially dependent actual concentration outside the droplets.
If the distance between droplets is large,
these features can be used to derive the generic laws of droplet ripening
in concentrations gradients and thereby
extend the classical  theory for homogeneous systems~\cite{Lifshitz_Slyozov_61,wagner61}.
Our central finding is that a regulator gradient leads to a drift velocity and
 a position dependent growth of drops (figure~\ref{fig:Fig_0}(b)). 
As a consequence, a dissolution boundary moves through the system, leaving droplets only in a region close to one boundary of the system. Using numerical calculations supported by analytic estimates, we study the growth dynamics of 
droplets in a gradient.
We discover that, surprisingly,  ripening is not always faster in the case of steeper regulator gradients. 
 Instead, a transient arrest of ripening is observed that results from
 a narrowing of the droplet size distribution. 
 Our work shows that a regulator gradient induces a novel and rich ripening dynamics in droplet systems.

\section{Local regulation of phase separation}

We use a simplified model to discuss two component phase separation 
that is influenced by a regulator.
We consider a system consisting of a solvent
$\Sol$, droplet material $\DD$ and a regulator $\RR$ \ch{that can create together with the droplet material a bound state $\RR$:$\DD$.}
 In this model the regulator does not take part in demixing but 
influences phase separation of $\DD$ and $\Sol$. 
We describe demixing by a \ch{simplified Flory-Huggins type of free-energy density}
\begin{equation}
\label{eq:FH_init}
	f= {k_B T} \bigg[ \frac{\phi^T_\DD}{\nu_{\DD}}  \ln{\phi^T_\DD} 
	+\frac{\phi_\Sol}{\nu_{\Sol}} \ln{\phi_\Sol}  \bigg]+ \mathcal{E} \, ,
\end{equation}
where $k_B$ is the Boltzmann constant, $T$ 
is temperature and  $\phi^T_\DD$ and $\phi_\Sol$ denote the total volume fraction of droplet material and solvent, respectively, 
with $\phi^T_\DD+ \phi_\Sol=1$. 
\ch{In equation~(\ref{eq:FH_init}) we neglect for simplicity the mixing entropy of the component 
$\phi_{\RR:\DD}$.
We only consider interactions between droplet  material $\DD$ and solvent $\Sol$ and the corresponding interaction energies are described by $\mathcal{E}$.
These simplifications do not affect the qualitative feature of position dependent phase separation that we highlight in this work but are useful simplification for the discussion of the relevant physics.}
The molecular volumes $\nu_i$ connect volume fractions with concentrations $c_i$
by $\phi_i=\nu_i c_i$. 
The regulator influences phase separation by binding to droplet material, 
 \begin{equation}
	 \DD+\RR\rightleftharpoons\RR:\DD \, .
\end{equation} 
 Here we consider the case where the bound state $\RR$:$\DD$ does not phase separate from the solvent. 
The total volume fraction of droplet material is given by the sum of contributions of
bound and free molecules, $\phi^T_\DD=\phi_\DD+\phi_{\RR:\DD} $.
The binding process between regulator and droplet material can be described 
by mass action with the equilibrium binding constant $K_0=c_{\RR:\DD}/(c_\DD c_\RR)$.
Using the simplification $\nu_{\RR:\DD}=\nu_\DD$, we write
$K_0=\phi_{\RR:\DD}/(\phi_\DD c_\RR)$.
  The interaction energy is given by $\mathcal{E}= {k_B T} \chi \, \phi_\DD  \phi_\Sol$, where $\chi$ 
is the interaction parameter. 
Expressing $\phi_\DD$ in terms of $\phi^T_\DD$ and considering a fast local equilibrium of the binding reaction we find
\ch{
\begin{equation}
	\mathcal{E}(\phi^T_\DD, \phi_\Sol)= {k_B T} \chi_\mathrm{eff} \,  \phi^T_\DD  \phi_\Sol \, ,
\end{equation}	 
with 
\begin{equation}
	\chi_\mathrm{eff}=\chi  \left( 1-  \frac{K\phi_\RR}{1+K  \phi_\RR} \right)
\end{equation}
 and $K=K_0/\nu_\RR $.
 The function $\chi_\mathrm{eff}$ describes the effective interaction between the solvent and the total droplet material which depends on the regulator.
 In the case of a vanishing regulator concentration, $\phi_\RR=0$, equation~(\ref{eq:FH_init}) reduces to the original Flory-Huggins model for \ch{binary polymer} blends~\cite{rubinstein2003polymer} with an interaction parameter $\chi_\mathrm{eff}=\chi$.
Increasing the concentration of the regulator leads to a decrease in 
the effective interaction parameter $\chi_\mathrm{eff}$.  
This decrease is more pronounced if the binding constant 
$K$ is larger, which amounts to more $\RR$ being bound to $\DD$.  
Please note that only for large values of $\chi_\mathrm{eff}$ \ch{relative to the entropic terms in equation~(\ref{eq:FH_init}) demixing can occur
(figure~\ref{fig:Fig_1}(a)).}}

\section{Spatial organization of phase separation} 

\begin{figure}[t]
\centering
\begin{tabular}{cc}
\includegraphics[width=0.4\textwidth]{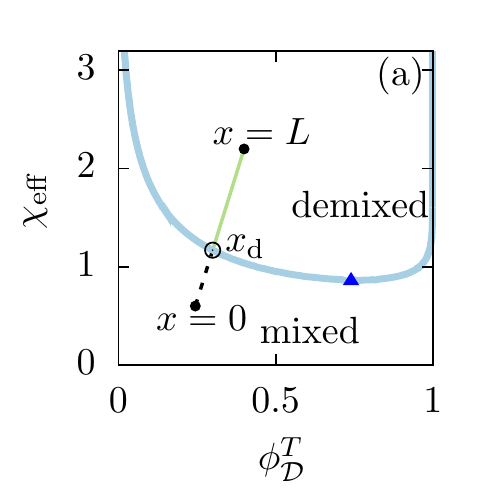} & 
\includegraphics[width=0.4\textwidth]{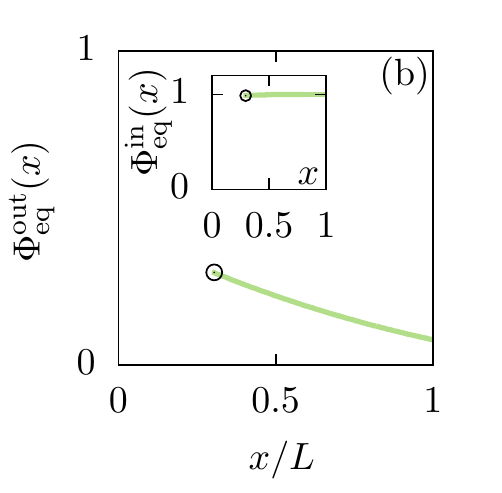} 
\end{tabular}
\caption{\label{fig:Fig_1} 
(a) Phase diagram. Interaction parameter $\chi_\mathrm{eff}$ as a function of the 
volume fraction of droplet material $\phi_\DD^T$. 
The binodal line (blue)  and the critical point (triangle) are indicated. 
For a regulator concentration gradient, the positions $x$ in the system are mapped to a line 
(dashed/green line in the mixed/demixed region). 
At the position $x=x_\mathrm{d}$ this line crosses the binodal.
(b) Equilibrium volume fractions outside and inside the droplet, $\phiout$ and $\phiin$, 
corresponding to the binodal line in (a), are shown as functions 
of position $x>x_\mathrm{d}$ for $K=500$. 
Parameters: $m=-3\cdot 10^{-3}$, $\phi_0=4\cdot 10^{-3}$, $\nu_\Sol =10 \nu_\DD$.}
\end{figure}

\ch{To describe the spatial regulation of phase separation} we consider a spatially inhomogeneous system that is
locally at thermodynamic equilibrium such that at each position the local 
free energy is defined. Globally the system is maintained away from equilibrium by an imposed position dependent regulator gradient.
For simplicity, we use a linear gradient along the $x$ direction, 
$\phi_\RR(x)=\phi_0-m\cdot x$,
with $x\in[0,L]$, where $L$ denotes the size of the system.
We first look at a situation 
without droplets but with a possible spatial profile $\phi_\DD=\phiinf(x)$ of droplet material.
Since the spatial concentration profile of the regulator $\phi_\RR(x)$ is imposed,
the effective interaction parameter $\chi_\mathrm{eff}(x)$ becomes a function of $x$. 
As the droplet material is also distributed in space, the concentration 
at each position $x$ corresponds to
a point in the phase diagram. The linear range $x\in[0,L]$ then maps onto a line 
that is indicated in the phase diagram in figure~\ref{fig:Fig_1}(a).
Using the phase diagram,  we can determine the position $x_\mathrm{d}$ of the dissolution
boundary, which separates the
region $x<x_\mathrm{d}$ where the fluid mixes, from the region $x>x_\mathrm{d}$ in which droplets can form. 
For $x>x_\mathrm{d}$, we can then determine the local equilibrium volume fraction $\phiin(x)$ of the droplet material inside  and $\phiout(x)$ outside of a potential droplet, which depend on position.
For  $\nu_\Sol \gg \nu_\DD$, $\phiin$ is approximately constant along $x$.
As we will see below, choosing this simple limit allows us 
to focus on the concentration field outside of the droplet.
The spatial distribution of the regulator and droplet material
imply a spatially dependent  supersaturation defined as
\begin{equation}\label{eq:supersaturation}
\epsilon(x)=\frac{\phiinf(x)}{\phiout(x)}-1 \, ,
\end{equation}
which is positive for $x>x_\mathrm{d}$. 
In the absence of droplets, the concentration field $\phiinf(x)$ evolves in time satisfying
a diffusion equation. If droplets are nucleated, their
 dynamics of growth or shrinkage is guided by the local supersaturation $\epsilon(x)$
 as well as  $\phiin$ and $\phiout(x)$.
 This droplet dynamics then in turn also influences the concentration field $\phiinf(x)$.

 \section{Dynamics of a single drop in a concentration gradient}

\ch{A regulator concentration gradient generates 
 a position-dependent supersaturation (equation~(\ref{eq:supersaturation})), 
 which  will generically influence the spatial distribution of droplet material $\phiinf(x)$.
 In the following we  discuss  the kinetics of growth of a single droplet where 
 the equilibrium concentration, $\phiout(x)$, and droplet material, $\phiinf(x)$,
 are position dependent, and thereby extend the classical description of droplet growth~\cite{Lifshitz_Slyozov_61,wagner61} to the case of concentration gradients.}
  
\ch{Neglecting variations of the concentration inside of the droplet, we 
restrict ourselves to the concentration field outside of the single droplet, $\phismallout(r,\theta,\varphi)$.
We use spherical coordinates centred at the droplet position $x_0$,
with  $r$ denoting the radial distance from the centre, and $\theta$ and $\varphi$ 
are the azimuthal and polar angles.
The volume fraction outside but near the droplet then obeys the steady state
of a diffusion equation 
\begin{equation}\label{eq:diff_equ}
	\nabla^2\phismallout(r,\theta,\varphi)=0 \, .
\end{equation}
The concentration field approaches for large $r$ (far from the drop) a linear gradient of the form, 
\begin{equation}\label{eq:bc1}
\lim_{r\to\infty} \phismallout(r,\ch{\theta})=\alpha \, r \cos \theta +\beta \, ,
\end{equation}
 where
  the droplet material outside $\phiinf(x)$ is locally characterized by the concentration  $\beta=\phiinf(x_0)$ and the gradient $\alpha=\partial_x\phiinf(x_0)$ at the droplet position $x_0$. 
  At the surface $r=R$ of a spherical droplet,
  the boundary condition is 
  \begin{equation}\label{eq:bc2}
  	\phismallout(R,\ch{\theta})=(\phiout + R \cos(\theta)\partial_x \phiout )(1+\ell_c/R) \, .
\end{equation} }
 Here, $\ell_\mathrm{c}= 2\gamma \nu_{D}/(k_bT)$ is the capillary length, 
 $\gamma$ denotes the surface tension of the droplet.
 \ch{Equation~(\ref{eq:bc2}) corresponds to the Gibbs-Thomson relation~\cite{Bray_Review_1994}, which
 describes the increase of the local concentration at the droplet interface relative to the equilibrium concentration due to the surface tension of the droplet.
The presence of  spatial inhomogeneities on the scale of the droplet $R$ lead to an additional term in the Gibbs-Thomson relation of the form $R \cos(\theta)\partial_x \phiout$.}
The values of $\alpha$ and $\beta$ characterizing the far field together with the local concentration at the droplet surface, $\phismallout(R,\ch{\theta})$, then determine the local rates of growth or shrinkage of the drop. 
Deformations of the spherical shape of the droplet can be neglected if the surface tension is large and concentration gradients on the scale of the droplet are small.
Furthermore, we focus, for simplicity, on the 
case where the Onsager cross coupling coefficient between the regulator and droplet material is negligible and we thus ignore
how the spatial distribution of droplet material affects the maintained regulator gradient.

  The solution to the diffusion equation~(\ref{eq:diff_equ})
  is  of the form, $ \phismallout (r,\theta) = \sum_{n=0}^\infty \left(A_n r^n +B_n r^{-n-1}\right) P_n (\cos \theta)$,
where $P_n (\cos \theta)$ are the Legendre polynomials. Using the boundary conditions~(\ref{eq:bc1}) and (\ref{eq:bc2}), we find
  \begin{eqnarray}
\nonumber
\phismallout(r, \theta)&= \alpha \cos \theta \left( r - \frac{R^3}{r^2} \right)
+\beta \left( 1- \frac{R}{r}\right) 
\\&+ \left(\phiout + R \cos(\theta)\partial_x \phiout \right) \left( 1+\frac{\ell_\mathrm{c}}{R}\right) \frac{R}{r} \, .
 \label{eq:phi_quasi_static}
\end{eqnarray}
\ch{The interface of }a droplet at position 
$x_0$ can be expressed by a function ${\mathpzc R}(\theta, \varphi, t; x_0)$. 
The 
\ch{speed of the interface} is $\partial_t {\mathpzc R}(\theta, \varphi,t; x_0)=v_\mathrm{n}(\theta, \varphi; x_0) $, 
where $v_\mathrm{n}=\vec n \cdot \vec{J}$ is the local velocity normal to the interface and $\vec n$ denotes a surface normal. 
Here, $\vec{J}=(\vec j^\mathrm{in}-\vec j^\mathrm{out})/(\Phi^\mathrm{in}_\DD-\phiout)$ is the 
\ch{local interface velocity}~\cite{Bray_Review_1994}, and
$\vec j^\mathrm{in}$ and $\vec j^\mathrm{out}$ denote the \ch{volume} 
 fluxes at the droplet surface inside and outside of the drop.  
Since the volume fraction inside the droplet is considered 
as constant and independent of the droplet position, 
$\vec j^\mathrm{in}=0$ and $\vec j^\mathrm{out}=-D \nabla \phismallout$.
 In the limit of strong phase separation ($\phiin \gg \phiout$)  the growth velocity normal to the interface is
$v_\mathrm{n}= (D/\phiin) \partial_r \phismallout|_{r=R} $.
 \ch{With the definition of the droplet radius, $R=(1/4\pi)\int \mathrm{d} \varphi \mathrm{d}\theta \sin\theta \, {\mathpzc R}$, 
 we can calculate the growth rate of the droplet radius, $\mathrm{d}R/\mathrm{d}t=(1/4\pi)\int \mathrm{d} \varphi \mathrm{d}\theta \sin\theta \, \partial_t {\mathpzc R}$,
and the net drift velocity along the $x$-direction,
${V}_\mathrm{d}=(1/4\pi)\int \mathrm{d} \varphi \mathrm{d}\theta \sin\theta \, \vec{e}_x \cdot \vec{e}_r \,  \partial_t {\mathpzc R}$. Here, 
$\vec{e}_x \cdot \vec{e}_r=\cos\theta$ and $\vec{e}_r$ and $\vec{e}_x$ denote the radial unit vector in spherical coordinates and the unit vector along  the $x$-direction in cartesian coordinates. 
Thus,  the droplet radius grows as
\begin{equation}\label{eq:LS_inhomog}
	\frac{ \mathrm{d}R}{\mathrm{d}t }=
	\frac{D}{\phiin R}\; \left[ \beta -\phiout(x_0) \left(1 + \frac{\ell_\mathrm{c}}{R} \right)\right] \, . 
\end{equation}
In the presence of concentration gradients there also exists a net drift velocity 
with
\begin{equation}\label{eq:drift}
	{V}_\mathrm{d}= \frac{D}{\phiin}  \left[ \alpha - \partial_x \phiout (x_0)  \left(1 + \frac{\ell_\mathrm{c}}{R} \right) \right] \, .
\end{equation}}
\ch{Note that both the growth speed and the drift velocity are set by the molecular diffusion constant $D$ of droplet material. }

 \section{Ripening of multiple drops in a regulator gradient}
 
 \ch{We can now describe the  dynamics of many droplets $i=1,\cdots, N$,
with positions $x_i$ and  radius 
$R_i$. If droplets are far apart from each other, the rate of growth of  droplet $i$ reads}
\begin{equation}\label{eq:LS_inhomogi}
	\frac{\mathrm{d} }{\mathrm{d}t} R_i=
	\frac{D}{R_i}\; \frac{\phiout(x_i)}{\phiin}  \left[ \epsilon(x_i) - \frac{\ell_\mathrm{c}}{R_i}\right] \, .
\end{equation}
The droplet drift velocity, $\mathrm{d} x_i/\mathrm{d}t= {V}_\mathrm{d}(x_i)$, is given by 
\begin{equation}\label{eq:drifti}
	\frac{\mathrm{d} x_i}{\mathrm{d}t}= \frac{D}{\phiin}  \left[ \partial_x  \phiinf(x)\vert_{x_i}  - \partial_x \phiout(x)\vert_{x_i}   \left(1 + \frac{\ell_\mathrm{c}}{R_i} \right) \right]  \, .
\end{equation}
If the distance between droplets is large relative to their size,
droplets only interact {\it via} the concentration field  $\bar \Phi(x,t)$ which represents the far field. 
It is governed by  a diffusion equation 
 including gain and loss terms associated with growth or shrinkage of drops:
\begin{equation}\label{eq:beta_eq}
	 \partial_t \bar \Phi(x,t) = D \frac{
	 	\partial^2}{
	 	\partial  x^2}  \bar \Phi(x,t) - \frac{4\pi \phiin}{3L^3} \sum^{N}_{i=1}\delta(x_i-x)\frac{\mathrm{d} }{\mathrm{d} t} R^3_i(t)
	  \, .
\end{equation}
For simplicity, in the above equation we consider a regulator gradient along the $x$ axis.
Please note that equation~(\ref{eq:beta_eq})
 describes the effects of large scale spatial inhomogeneities on the ripening dynamics. Since large scale variations of $\bar \Phi(x,t)$ only build up along the $x$-directions, derivatives of $\bar \Phi$ along the $y$ and $z$
 axes do not contribute.

\ch{In the absence of a regulator gradient, $\phiout$ and  $\bar \Phi$ are constant
implying a position-independent supersaturation level $\epsilon$ (equation~(\ref{eq:supersaturation})).
In this case equation~(\ref{eq:LS_inhomogi})  gives the  classical 
law of droplet ripening derived by Lifschitz-Slyozov ~\cite{Lifshitz_Slyozov_61,wagner61} (also referred to as Ostwald-ripening),  and the net drift vanishes  (equation~(\ref{eq:drifti})).
In the case of Ostwald ripening large droplets of radius larger than the critical radius,  $R_{\mathrm{c}}= \ell_{\mathrm{c}}/\epsilon$, grow at the expense of smaller shrinking drops.
This  causes an increase of the average droplet size and a broadening of the droplet size distribution with time. On large spatial scales, droplets remain homogeneously distributed in the system.}

\ch{This property fundamentally changes due to the presence of concentration gradients leading to  two  possibilities of droplet material transport along the regulator gradient: 
(i) Exchange of material between droplets at different positions of the concentration gradient by diffusive transport in the dilute phase or
 (ii) drift of  droplets along the concentration gradient. 
(i) Droplets grow or shrink with rates that vary along the gradients of  local equilibrium volume fraction $\phiout$ and the droplet material volume fraction $\bar \Phi(x)$ (Eq.~(\ref{eq:LS_inhomogi})). 
 For $\epsilon(x)=\phiinf(x)/\phiout(x)-1 > \ell_\mathrm{c}/R$, a droplet located at position $x$ grows, and shrinks in the opposite case. The critical droplet radius thus  becomes  position dependent, where  below or above $R_{\mathrm{c}}(x)= \ell_{\mathrm{c}}/\epsilon(x)$  droplets shrink or grow.
(ii) The drift of a droplet (equation~(\ref{eq:drifti})) results from 
an asymmetry of material flux though the interface parallel to the regulator gradient.
If $|\partial_x  \phiinf(x)|< |\partial_x \phiout|$,  the droplet drift velocity ${V}_\mathrm{d}$ 
 points toward  regions of smaller $\phiout(x)$. 
This is a typical case since the gradient of droplet material
$ \partial_x  \phiinf(x)$ {\ch tends to flatten with time due to the diffusion of droplet material in the dilute phase.}}

%
%
\begin{figure}[t]
\centering
\begin{tabular}{cc}
\includegraphics[width=0.4\textwidth]{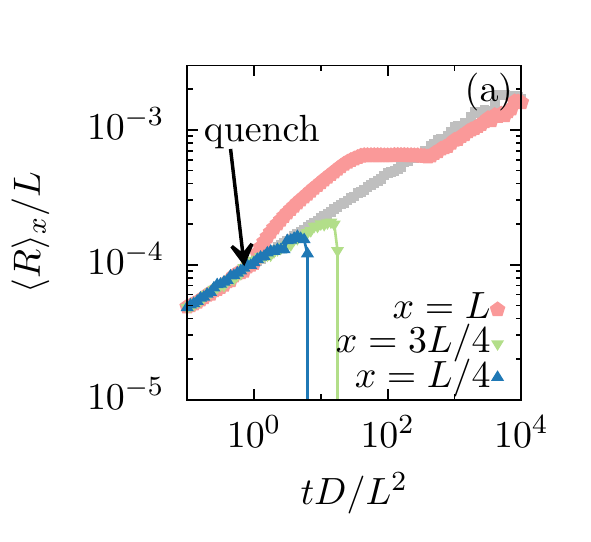} 
\includegraphics[width=0.4\textwidth]{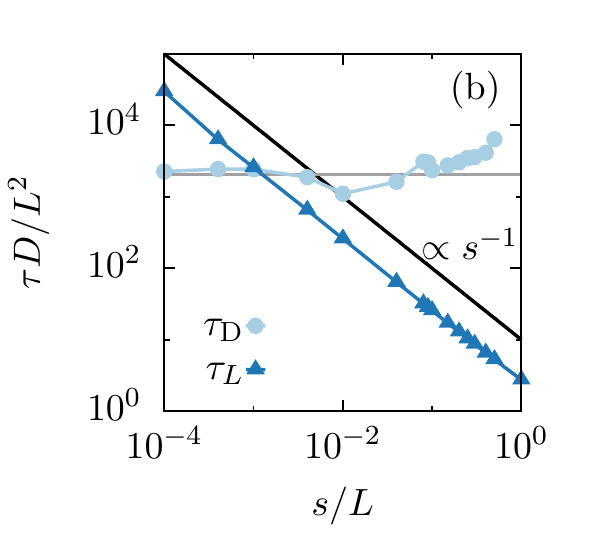}
\end{tabular}
\caption{\label{fig:Fig_2} 
Droplet ripening in concentration gradients.
(a) Mean droplet radius $\langle R \rangle_x$ at position $x$ as a function of time for different $x$
as indicated.  A spatial profile of equilibrium volume fraction of slope $s=0.5$ is imposed
at time $t=L^2/D$ (quench). \ch{The grey data points correspond to classical Ostwald ripening ($s=0$).}
(b) Characteristic time $\tau_L$ required to segregate the volume of droplet material 
toward $x=L$, and dissolution time $\tau_\mathrm{D}$ required to reach $10$ droplets starting from $\mathcal{O}(10^4)$, 
as a function of quench slope $s$. The horizontal grey line indicated the value of
$\tau_\mathrm{D}$ for classical Ostwald ripening ($s=0$).
}
\end{figure}
%
 \begin{figure}[t]
\centering
\begin{tabular}{cc}
\includegraphics[width=0.4\textwidth]{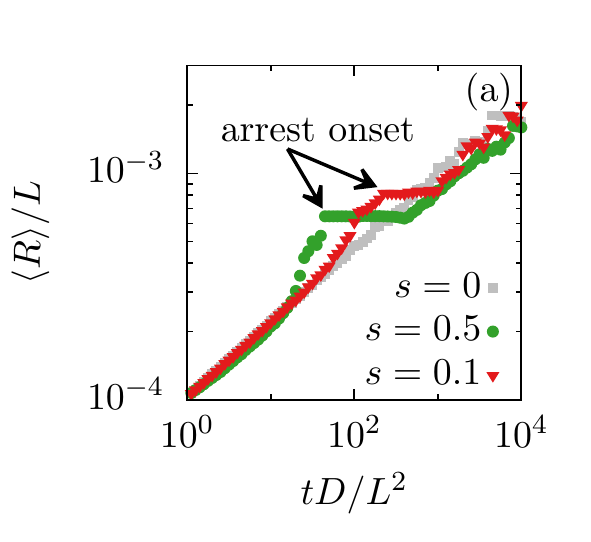} &
\includegraphics[width=0.4\textwidth]{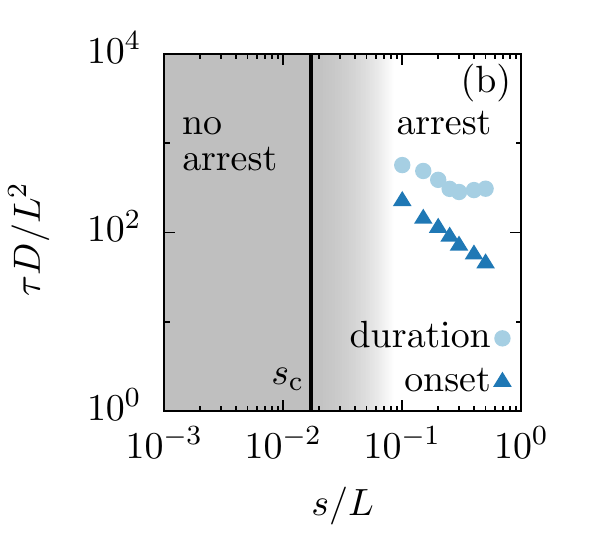}\\
\includegraphics[width=0.4\textwidth]{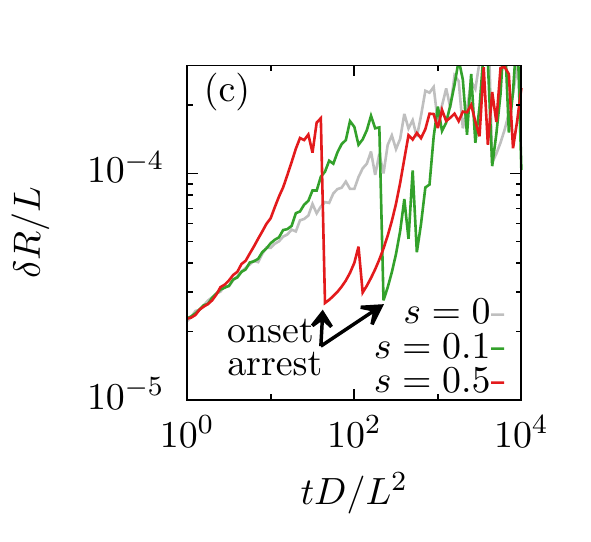} &
\includegraphics[width=0.4\textwidth]{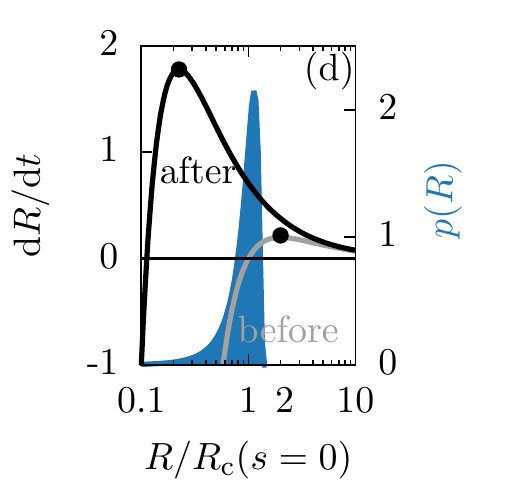}  
\end{tabular}
\caption{\label{fig:Fig_3} 
Narrowing droplet size distribution.
(a)  Mean radius $\langle R \rangle$ averaged over all drops in the system as a function of time for three quench slopes $s$. The onset of arrest is indicated (arrows).
(b) Duration and time of onset of the arrest phase as a function of quench slope $s$.  
The vertical black line indicates the quench slope $s_\mathrm{c}=\{1-[3/2-\Phi_0/(2\bar \Phi(t=L^2/D))]^{-1}\}/L$ below which no arrest can occur.
It can be calculated by the condition that the critical radius $R_\mathrm{c}$ at $x=L$ is reduced by at least a factor of $3/2$ during the quench such that the largest droplets in the distribution grow more slowly than smaller ones and the distribution narrows.
For our numerical solutions $s_\mathrm{c}\approx0.017/L$, which is consistent with the emergence of the arrest along quench slope found in our numerical calculations.
(c) Standard deviation $\delta R$ of the droplet 
radius distribution as a function of time for three different quench slopes $s$. The onset of arrest corresponds to a sudden narrowing
of the distribution (arrows).
(d) Rate of droplet growth $\mathrm{d} R/\mathrm{d}t$ as a function of droplet
radius $R$ before (grey) and after (black) the spatial quench. The
droplet radius distribution $p(R)$ 
at the moment of the quench is shown (blue).
Narrowing of $p(R)$ occurs if droplet size
exceeds the radius for which the 
growth rate is maximal (black dots). }
\end{figure}
%
To study the ripening dynamics of droplets in a concentration gradient
we solved the equations~(\ref{eq:LS_inhomogi}) to (\ref{eq:beta_eq}) numerically.
 To access the late time regime of ripening we first initialize about
 $N=10^7$ drops 
 with radii taken from the Lifschitz-Slyozov distribution~\cite{Lifshitz_Slyozov_61,wagner61} in a system of
 position independent equilibrium concentration $\Phi_0$, and fix the concetration inside $ \phiin=1$. 
 For $t\ge L^2/D$, we then spatially quench  the system by imposing 
 the  spatially varying equilibrium concentration $\phiout(x)= \Phi_0  (1- s\, x)$~\cite{footnote2},
which we refer to as ``spatial quench'' in the following.
In our numerical studies we find that droplets experience a non-uniform growth depending on the position and the stage of ripening (figure~\ref{fig:Fig_2}(a)). 
At the beginning, all drops grow in the region where the concentration 
$\phiinf(x)$ exceeds the local equilibrium concentration at the drop surface,
$\phiout(x) \left(1+\ell_\mathrm{c}/R\right)$, and shrink otherwise.   The dissolution boundary at 
$x=x_\mathrm{d}$
obeys \ch{$\bar\Phi(x_\mathrm{d}) \simeq \phiout (x_\mathrm{d})$ since in the late time regime $\ell_\mathrm{c} \ll R$}.  It moves according to 
\begin{equation}
	\frac{\mathrm{d} x_\mathrm{d\mathrm}}{\mathrm{d}t} = {\frac{\mathrm{d} \bar \Phi(x_\mathrm{d}(t))} {\mathrm{d}t}  }
	\bigg/{  \left.\frac{ \mathrm{d}\phiout (x)}{\mathrm{d}x}\right|_{x=x_\mathrm{d}(t)} } \, .
\end{equation}
For $\mathrm{d}\phiout (x)/\mathrm{d}x <0$, the position of the dissolution boundary
$x_\mathrm{d}$ 
moves to the right until it reaches the system boundary at $x= L$ (Supplemental video~\cite{SM}).
At long times, the volume fraction at all positions  
approaches the minimum of the equilibrium volume fraction, $\bar\Phi(x)\to \phiout(L)$, 
and all droplets dissolve except at $x=L$ (figure~\ref{fig:Fig_2}(a), Supplemental video~\cite{SM}).

The characteristic time $\tau_L$ of droplet segregation depends on the  quench slope $s$.
 It decreases for increasing $s$ according 
 $\tau_L\propto s^{-1}$ (figure~\ref{fig:Fig_2}(b)).
 In contrast the time of droplet dissolution, $\tau_\mathrm{D}$, defined as the time to reach 10 droplets,  changes only weakly with the quench slope $s$ and can even increase  (figure~\ref{fig:Fig_2}(b)). Interestingly, the droplet ripening exhibits periods of transient arrest, during which droplet number and size remain almost constant (figure~\ref{fig:Fig_3}(a)).
 These arrest phases govern the time of droplet dissolution for large quench slopes since they occur for sufficiently large quench slopes $s$.
The duration of arrest is roughly constant as a function of $s$
and the onset of the arrest phase is delayed for decreasing $s$~\cite{footnote1} (figure~\ref{fig:Fig_3}(b)). 
\ch{Intriguingly, the onset of the arrest phase is preceded by a 
narrowing of the droplet size distribution.
The droplet size distribution narrows during the segregation of droplets toward $x=L$
 while the onset of arrest occurs after droplets have mostly been spatially segregated (Supplemental video~\cite{SM}).}
In particular, the standard deviation of droplet radius
exhibits a pronounced minimum when the arrest begins  (figure~\ref{fig:Fig_3}(c)). 
After the arrest phase droplets undergo classical Ostwald ripening where time-dependence of
  $\langle R \rangle$ and $\delta R$  is consistent with $t^{1/3}$
(figures~\ref{fig:Fig_2}(a) and \ref{fig:Fig_3}(a)). 
The effect of a narrowing droplet size distribution has also been observed 
in \ch{open but} spatially homogeneous systems with constant influx of phase separation 
material~\cite{clark_nanolett11,Vollmer_size_focussing}.

The narrowing  of the droplet size distribution in a concentration gradient 
is fundamentally 
different from  broadening of the droplet size distributions during
classical Ostwald-ripening~\cite{Lifshitz_Slyozov_61,wagner61}.
 Ostwald ripening is characterized by a 
 supersaturation that decreases with time, leading to an increase of the  critical droplet radius  $R_\mathrm{c}=\ell_\mathrm{c}/\epsilon(t)\propto t^{1/3}$. The droplet size distribution $p(R)$
 has a universal shape and is nonzero only in the interval $[0, 3R_\mathrm{c}/2]$ (figure~\ref{fig:Fig_3}(d), \ch{blue graph}).
 The broadening of $p(R)$ follows from larger droplets growing at a larger rate $\mathrm{d}R/\mathrm{d}t$ than smaller droplets.
  Though $\mathrm{d}R/\mathrm{d}t$ has a
 maximum at $R=2R_\mathrm{c}$ 
and decreases for large $R$, no droplets exist larger than $3R_\mathrm{c}/2$.

This situation changes in the presence of a concentration gradient.
The spatial quench reduces the local critical radius $R_\mathrm{c}(x\simeq L)=\ell_\mathrm{c}/\epsilon(x\simeq L)$ 
at the rightmost boundary  $x\simeq L$ as compared to the critical radius before the quench (equation~(\ref{eq:supersaturation})). 
This quench also shifts the maximum of $\mathrm{d}R/\mathrm{d}t$ for droplets
 at $x\simeq L$ to smaller radii (black line in figure~\ref{fig:Fig_3}(d)) since  the  radius corresponding to the maximum occurs at $R=2 R_\mathrm{c}$. 
As a result, many droplets now exist after the spatial quench with large 
radii  $R>2 R_\mathrm{c}(x\simeq L)$.
These droplets grow more slowly than those at $R=2R_\mathrm{c}$ which
leads to a narrowing of the size distribution $p(R)$ at $x\simeq L$. 
 The critical radius $R_\mathrm{c} (x \simeq L)$ remains small because dissolution of droplets at $x<L$ leads to a diffusive flux toward $x\simeq L$ and thus keeps the volume fraction $\bar \Phi(L)$ at increased levels.
 These conditions hold longer if the spatial quench has a steeper slope. As a result the distribution  narrows more for steeper quenches.
 When the critical radius catches up with the mean droplet size narrowing stops and the onset of arrest occurs.
 At this time droplets have almost equal size which 
  slows down the exchange of material between droplets {\it via} Ostwald ripening, leading to a long phase of almost constant size and number of droplets (figure~\ref{fig:Fig_2}(a)). 
  During this arrest phase, the droplet distribution broadens slowly.

\section{Conclusion and Outlook}

\ch{Here we presented the generic behavior of droplet ripening
in concentrations gradients and 
extended the classical  theory by Lifschitz \& Slyozov to inhomogeneous systems~\cite{Lifshitz_Slyozov_61,wagner61}.}
\ch{One main result is that a concentration gradient of a soluble  component that regulates liquid-liquid phase separation 
can reshape the supersaturation profile such that all drops dissolve except those within a region close to one boundary of the system. As a consequence droplets segregate toward the boundary where the supersaturation is highest~\cite{Lee_2013}.
Even though the details by which a regulator affects
the local supersaturation are system-specific, the resulting
ripening dynamics that takes place in a supersaturation gradient is generic.
Surprisingly, we find that the size distribution of droplets  narrows for sufficiently steep concentration gradients, 
leading to a transient arrest of the droplet dynamics.
Such a behavior is fundamentally different to classical Ostwald-ripening where the droplet size distribution continuously broadens at all times.
Transient narrowing of the droplet size distribution stems from a position-dependent shift of the maximal droplet growth rate to smaller droplet radii as compared to spatially homogeneous systems (figure~\ref{fig:Fig_3}(d)).
}

\ch{Our work shows that droplet ripening in concentration gradients exhibits fundamental differences compared to classical phase-separating systems where droplet positions are homogeneously distributed in space. 
The physics presented here could be relevant for the control of emulsions in chemical engineering and biology.
The narrowing of droplet size distributions found in the presence of a regulator gradient could be used to control droplet size in emulsions. It  provides a physical mechanism for the formation of almost mono-disperse emulsions.
An example in biology where an emulsion is controlled by concentration gradients is the  \emph{C.elegans} embryo~\cite{Brangwynne_2009, Lee_2013, saha2016polar}. 
 In this system liquid-like cellular compartments, so called P granules, are positioned toward the posterior side of the cell prior to asymmetric cell division by a protein concentration gradient.
 An increasing number of membrane-less compartments with liquid-like properties have been characterized~\cite{hyman2014liquid,Hyman_Alberti}. 
Their formation and positioning  could be a general scheme for the spatial organization of chemistry in living cells.
  In our work we have identified the physical mechanisms of spatial segregation of droplets by concentration gradients.
  The physics discussed here contributes to the behavior of liquid-like compartments in living cells such as P granules.
  However, many aspects of the dynamics of liquid-like compartments inside cells  remain unexplored.
 In particular, they consist of a large number of components  and are chemically active. 
  Emulsions in the presence of chemical reactions driven away from equilibrium can give rise to novel phenomena in phase separating systems such as the suppression of Ostwald ripening~\cite{Zwicker_Suppression_Ostwald} or the spontaneous division of  liquid droplets~\cite{Zwicker_active_drops}.
 Future questions could address how  nucleation, fusion, and droplet shape 
 is changed by concentration gradients 
 and how non-equilibrium chemical reactions in droplets are affected  by concentration gradients.}

\ack
We would like to thank Shambaditya Saha  and Anthony A.\ Hyman for stimulating discussions. 

\section*{References}


\begin{thebibliography}{10}
\bibitem{footnote2}
We checked that a quench of the form $\phiout(x)= \Phi_0 [1- s (x-L/2)]$ leads
  to qualitatively similar results;.
\bibitem{footnote1}
The arrest phase is defined as the time interval during which $N(t)$ decreases
  more slowly then $t^{-1/2}$ For Ostwald ripening ($s=0$) $N(t)\sim t^{-1}$;.
\end{thebibliography}

\end{document}